\documentclass[journal]{IEEEtran}
\usepackage[T1]{fontenc}
\usepackage[latin9]{inputenc}
\usepackage{amsmath}
\usepackage{amssymb}
\usepackage{amsthm,bm}
\usepackage{url}
\usepackage{cite}
\usepackage{babel,verbatim,balance}
\usepackage[rgb]{xcolor}
\usepackage{graphics, graphicx}
\usepackage{acronym}
\usepackage{upgreek}
\usepackage[bookmarks,colorlinks]{hyperref}
\usepackage{setspace,gensymb}

\usepackage{float}
\usepackage{pgfplots}
\usepackage[bookmarks,colorlinks]{hyperref}
\usepackage[linesnumbered,ruled,lined]{algorithm2e}
\usepackage{enumitem}
\usepackage{algpseudocode}
\usetikzlibrary{shapes.multipart,intersections}
\usepackage{cite}
\usepackage{amsmath,amssymb,amsfonts,amsthm,steinmetz}
\usepackage{graphicx}
\usepackage{mathrsfs}  
\usepackage{textcomp}
\usepackage{acronym}
\usepackage{xcolor}
\usepackage{upgreek,xspace}

\usepackage{tikz}
\usetikzlibrary{calc}
\makeatletter
\newcommand{\gettikzxy}[3]{%
  \tikz@scan@one@point\pgfutil@firstofone#1\relax
  \edef#2{\the\pgf@x}%
  \edef#3{\the\pgf@y}%
}

\begin{document}

\title{Mutual Coupling in Dynamic Metasurface Antennas: Foe, but also Friend}

\author{Hugo Prod'homme and Philipp del Hougne
\thanks{Hugo~Prod'homme and Philipp~del~Hougne are with Univ Rennes, CNRS, IETR - UMR 6164, F-35000, Rennes, France.
}
\thanks{\textit{Corresponding Author: Philipp del Hougne (e-mail: philipp.del-hougne@univ-rennes.fr).}}
}

\maketitle
\begin{abstract}
Dynamic metasurface antennas (DMAs), surfaces patterned with reconfigurable metamaterial elements (meta-atoms) that couple waves from waveguides or cavities to free space, are a promising technology to realize 6G wireless base stations and access points with low cost and power consumption. Mutual coupling between the DMA's meta-atoms results in a non-linear dependence of the radiation pattern on the DMA configuration, significantly complicating modeling and optimization. Therefore, mutual coupling has to date been considered a vexing nuance that is frequently neglected in theoretical studies and deliberately mitigated in experimental prototypes. Here, we demonstrate the overlooked property of mutual coupling to boost the control over the DMA's radiation pattern. Based on a physics-compliant DMA model, we demonstrate that the radiation pattern's sensitivity to the DMA configuration significantly depends on the mutual coupling strength. We further evidence how the enhanced sensitivity under strong mutual coupling translates into a higher fidelity in radiation pattern synthesis, benefiting applications ranging from dynamic beamforming to end-to-end optimized sensing and imaging. Our insights suggest that DMA design should be fundamentally rethought to embrace the benefits of mutual coupling. We also discuss ensuing future research directions related to the frugal characterization of DMAs based on compact physics-compliant models.
\end{abstract}

\section{Introduction}

Base stations and access points are expected to connect to massive numbers of user equipments under stringent cost and efficiency requirements in the 6th generation (6G) of wireless communications networks. Achieving these goals with conventional antenna arrays would involve very large numbers of radio-frequency (RF) chains and result in substantial cost, footprint, weight and power consumption, potentially undermining the realizability and scalability. A promising emerging technology poised to overcome these challenges are dynamic metasurface antennas (DMAs)~\cite{huang2020holographic,shlezinger2021dynamic}. A DMA is a thin (planar or conformal) surface patterned with tunable metamaterial elements (referred to as meta-atoms in this paper) that couple waves from waveguides or cavities to free space. The DMA's radiation pattern is hence the superposition of the fields radiated by its meta-atoms. The field radiated by a given meta-atom depends on its configuration as well as its excitation via the waveguide or cavity into whose surface it is embedded. Typically, the number of meta-atoms largely exceeds the number of feeds that excite the waveguides or cavities (each feed requires a costly RF chain). Therefore, DMAs can be interpreted as hybrid analog/digital architectures in which the transmitted or received signals are naturally processed in the analog domain as the waves propagate within the DMA, without additional dedicated analog combining architectures~\cite{shlezinger2021dynamic}. Optimally leveraging these advanced analog signal processing capabilities for maximally flexible and compact, low-cost, energy-efficient next-generation transceivers requires a rigorous understanding of the underlying wave phenomena.

The crux lies in how the excitation of a given meta-atom depends on the other meta-atoms' configurations. In general, a wave must be expected to bounce back and forth between the meta-atoms, implying that the excitation of a given meta-atom depends on how the other meta-atoms are configured. Ultimately, this mutual coupling between the meta-atoms results in a non-linear mapping from the meta-atoms' configurations (referred to as DMA configuration in this paper) to the resulting radiation pattern. Such a non-linear forward model severely complicates the analysis, characterization, optimization and deployment of DMAs. Therefore, most existing studies on DMAs have assumed negligible mutual coupling between the meta-atoms~\cite{shlezinger2021dynamic,smith2017analysis,boyarsky2021electronically,HISC}. Under this assumption, the excitation of a given meta-atom is independent from the configuration of the other meta-atoms, implying a linear (or, more precisely, affine) mapping from DMA configuration to radiation pattern. To ensure the validity of this assumption, early experimental prototypes have deliberately mitigated mutual coupling between meta-atoms by choosing an embodiment of the DMA concept based on stacked microstrips that are sparsely patterned with meta-atoms~\cite{boyarsky2021electronically}. Meanwhile, alternative embodiments based on quasi-2D chaotic cavities~\cite{sleasman2020implementation} that involve significant mutual coupling between meta-atoms do not appear in the wireless communications literature to date, presumably because modeling the strongly non-linear mapping from their configuration to the corresponding radiation pattern is perceived to be prohibitively complicated. Overall, it thus appears that mutual coupling is a vexing nuance that can and should be mitigated to simplify the deployment of DMAs.

Here, we challenge this exclusively negative perception of mutual coupling in DMAs by establishing an overlooked benefit of mutual coupling in DMAs. Indeed, we demonstrate in this paper based on a rigorous physics-compliant analysis that mutual coupling between meta-atoms boosts the degree of control over the DMA's radiation pattern. Intuitively, this can be understood as follows: The more a wave bounces back and forth between the meta-atoms, the more sensitivity to the DMA configuration it accumulates, which manifests itself not only as a stronger non-linearity in the forward model but also as enhanced control over the radiation pattern. Ultimately, to achieve maximal radiation pattern control with the most compact DMA structure and using the lowest possible number of meta-atoms, deliberately enhancing mutual coupling may turn out to be an appealing route -- diametrically opposed to the current trend of mitigating and/or neglecting mutual coupling in DMAs.

In this article, we begin by providing a high-level implementation-independent overview of physics-consistent DMA modeling which we then specialize to chaotic-cavity-backed DMAs that are ideally suited to systematically study the impact of the mutual coupling strength. We go on to demonstrate that the sensitivity of the radiated field to the DMA configuration increases as the mutual coupling between the meta-atoms gets stronger. Then, we evidence that the observed sensitivity increase manifests itself in terms of a higher fidelity with which desired radiation patterns can be synthesized. 
We conclude by looking forward to key research challenges and open questions in terms of DMA design and deployment that ensue from attempting to reap the benefits of mutual coupling highlighted in this article.

\section{Physics-Consistent DMA Forward Model}
\label{sec_forwardModel}

\subsection{High-level overview}

An important prerequisite for tractable physics-based modeling of DMAs is that the meta-atoms are lumped radiators. Indeed, meta-atoms are usually very small compared to the wavelength such that they are amenable to a description in terms of point-like dipoles. This facilitates the modeling of wave propagation both within the DMA as well as from the DMA to the region of interest (ROI). The importance of physics-based models lies in their compactness and favorable inductive bias, as discussed in more detail in Sec.~\ref{sec_challenges}. 

A dipole can be understood as a polarizable particle that accumulates a dipole moment upon illumination with an electromagnetic wave. The dipole moment is a charge separation across the particle's very small size that oscillates at the frequency of the impinging wave. The tendency of a dipole to acquire a dipole moment upon illumination is characterized by the dipole's polarizability (which is tunable in the case of the meta-atoms). A dipole with oscillating dipole moment will radiate in turn an electromagnetic field (again at the same frequency). The field at some position of interest originating from a unit dipole moment is known as the Green's function between the location of the dipole and the position of interest. The Green's function depends on the so-called ``background scattering'' in the propagation environment. For the special case of free space, the Green's function is known in closed form. In problems involving multiple dipoles, a given dipole's radiation induces secondary dipole moments in the other dipoles, which then in turn induce secondary dipole moments in the given dipole, etc. This multiple scattering of a wave between the dipoles gives rise to the mutual coupling phenomenon that physics-consistent models seek to capture.

Irrespective of the detailed DMA architecture, the generic approach to physics-consistent DMA modeling consists of five steps which are summarized as follows:
\begin{enumerate}
    \item Associate one dipole with each feed and each meta-atom.
    \item Determine the background Green's functions via which the dipoles are coupled to each other.
    \item Assemble the system's interaction matrix.
    \item Invert the system's interaction matrix to identify the meta-atoms' dipole moments.
    \item Propagate the fields radiated by each meta-atom's dipole moment to the ROI.
\end{enumerate}
The specific values of the background Green's function will depend on the considered DMA architecture, which will determine the ease with which they can be evaluated for step 2 in closed form, numerically or experimentally. Regardless, the above description is valid. If the DMA architecture has been engineered to decouple certain meta-atoms, then the values of the corresponding Green's functions will be low (ideally zero). Meanwhile, assuming the DMA is surrounded by free space, step 5 is simply a linear projection of the meta-atoms' dipole moments that is known in closed form.

The details of how to assemble the interaction matrix in step 3 can be found in relevant technical papers such as~\cite{del2020learned} and references therein. It is helpful to understand the role of the matrix inversion in step 4 which captures the mutual coupling phenomenon in a self-consistent manner. To conceptualize a clear physical picture, note that the matrix inversion can be cast into a converging infinite sum of matrix powers; the $k$th term of this sum physically corresponds to paths that bounce $k$ times between the dipoles~\cite{rabault2024tacit}. Longer paths are attenuated more strongly which is why the series converges. In DMAs with negligible mutual coupling, the higher-order terms in the series are negligible, giving rise to an effectively linear dependence of the radiation pattern on the DMA configuration. In contrast, in DMAs with strong mutual coupling, the wave bounces back and forth between the dipoles many times, implying that higher-order terms in the series are significant and the series converges more slowly. On the one hand, this implies a stronger non-linearity of the forward mapping from DMA configuration to radiation pattern, which explains why this regime has been avoided to date. On the other hand, the wave accumulates more sensitivity to the DMA configuration, which is an overlooked benefit of mutual coupling that we highlight in this article.

\subsection{DMA embodiments}

\begin{figure*}[t]
\centering
\includegraphics [width = \linewidth]{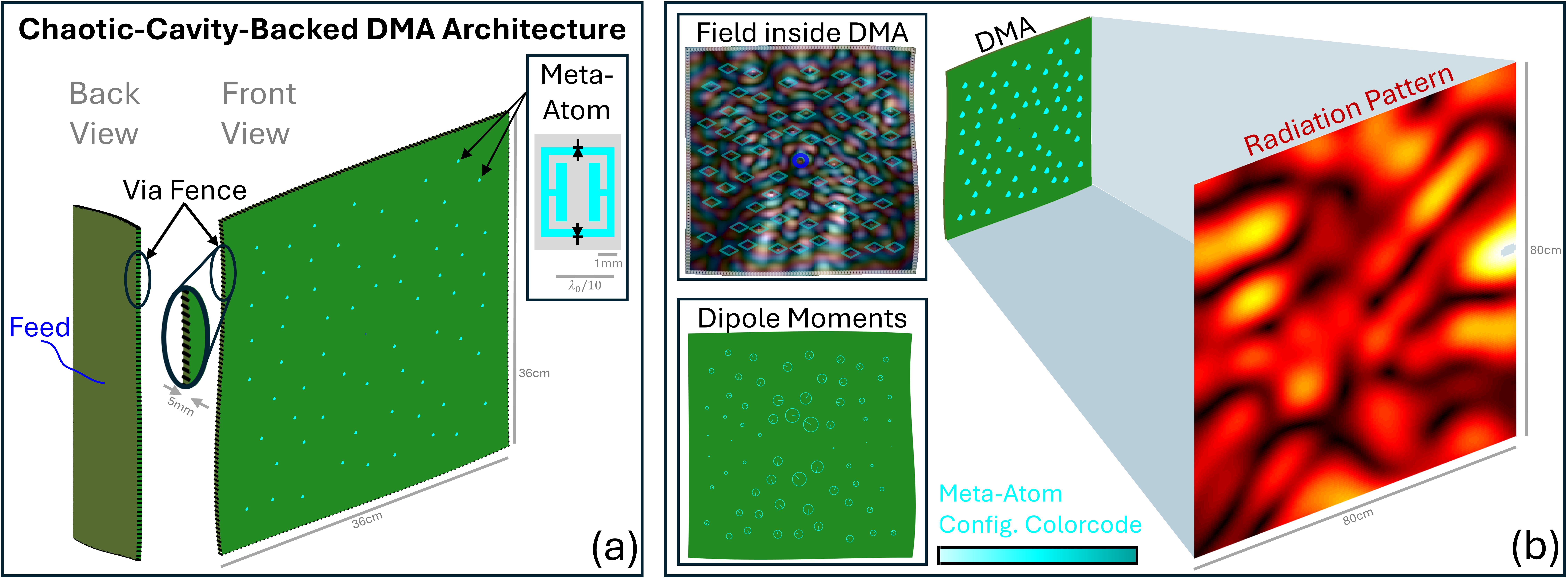}
\caption{\textbf{Architecture and working principle of chaotic-cavity-backed DMA (CCB-DMA).} (a) The considered CCB-DMA consists of a quasi-2D cavity made up of two parallel conducting plates and a via fence. It is excited by a single feed from the back; 64 randomly placed reconfigurable meta-atoms at the front leak energy from the cavity to the far field. The meta-atoms are continuously tunable cELC resonators parametrized by varactor diodes (see~\cite{sleasman2020implementation} for details). The via fence has an irregular shape to induce wave chaos. (b) The radiation pattern magnitude at 10~GHz observed 1m in front of the DMA is illustrated for a non-optimized uniform DMA configuration. The corresponding field distribution inside the DMA and the meta-atoms' dipole moments are provided in the insets; magnitude and phase of the complex-valued field distribution are represented by luminance and hue along a constant chroma in the HCL color space, respectively. }
\label{fig1}
\end{figure*}

The DMA embodiment based on stacked microstrips patterned with tunable meta-atoms is the only one currently considered in the wireless communications literature; a corresponding physics-compliant model capturing mutual coupling can be found in~\cite{williams2022electromagnetic}. Meanwhile, efforts to formulate physics-compliant models also exist for DMA embodiments based on a parallel plate waveguides or 2D chaotic cavities (see~\cite{del2020learned,yoo2020analytic} and references therein) in which mutual coupling between the meta-atoms is strong and not avoidable. However, these works are only concerned with formulating such physical models, and using them for optimization in the case of~\cite{del2020learned}. A discussion of potential benefits of mutual coupling is not provided, and the difficulties resulting from the non-linearity of the forward model are only implicit. Given these difficulties, it remains unclear what might motivate the consideration of such systems for technological applications. By unveiling a significant overlooked benefit of mutual coupling, our present paper answers this question.

To assess the impact of the mutual coupling strength on the radiation pattern sensitivity and synthesis, it is helpful to consider a DMA embodiment in which it is easy to vary the mutual coupling strength systematically. For this purpose, the chaotic-cavity-backed DMA (CCB-DMA) embodiment proposed in~\cite{sleasman2020implementation} is ideally suited. 
We stress, however, that our general conclusions are not restricted to this specific DMA embodiment. In the CCB-DMA, the feed(s) excite(s) a quasi-2D irregularly shaped cavity into whose upper surface the meta-atoms are patterned. In the printed-circuit-board realization of~\cite{sleasman2020implementation} illustrated in Fig.~\ref{fig1}(a), the cavity is implemented with a vertical via fence sandwiched between two parallel conducting plates. The in-plane dimensions of the cavity are on the order of 10 wavelengths whereas its height (the separation between the two conducting plates) is small compared to the wavelength. The cavity is hence effectively two-dimensional. The meta-atoms are electrically small, complimentary electric-LC (cELC) resonators etched into the upper conducting plane -- see inset in Fig.~\ref{fig1}(a); each meta-atom's polarizability is individually tunable in a continuous manner by adjusting the bias voltage of embedded varactor diodes. 
The wave reverberates within the in-plane cavity, scattering off the meta-atoms and the cavity boundaries defined by the vias. By adjusting the density of the via fence, the cavity's reverberation strength can be conveniently adjusted. Thereby, we can systematically vary the strength of the multiple scattering between the meta-atoms and hence their mutual coupling. 
The meta-atoms leak part of the wave energy incident on them out of the in-plane cavity by radiating it into free space. A meta-atom's in-plane scattering and out-of-plane leakage depend on its tunable polarizability. 
Details on how to analytically evaluate the background Green's functions of the CCB-DMA, how to estimate the meta-atom polarizabilities, and how to calculate the radiation pattern in the ROI can be found in technical papers like~\cite{del2020learned,yoo2020analytic} and references therein.

\section{Radiation Pattern Sensitivity \\to DMA Configuration}

We begin our analysis of the influence of the mutual coupling strength by considering the radiation pattern sensitivity to the DMA configuration. Clearly, the higher this sensitivity is, the more control the DMA configuration is expected to offer over the radiation pattern, and indeed we confirm this in the susbsequent Sec.~\ref{sec_synthesis}. The radiation pattern sensitivity is defined as the partial derivative of the radiation pattern with respect to the DMA configuration, which can be conveniently evaluated by implementing the differentiable forward model sketched in the previous Sec.~\ref{sec_forwardModel} in an automatic differentiation framework (e.g., JAX, TensorFlow, PyTorch).

\begin{figure}
\centering
\includegraphics [width = \linewidth]{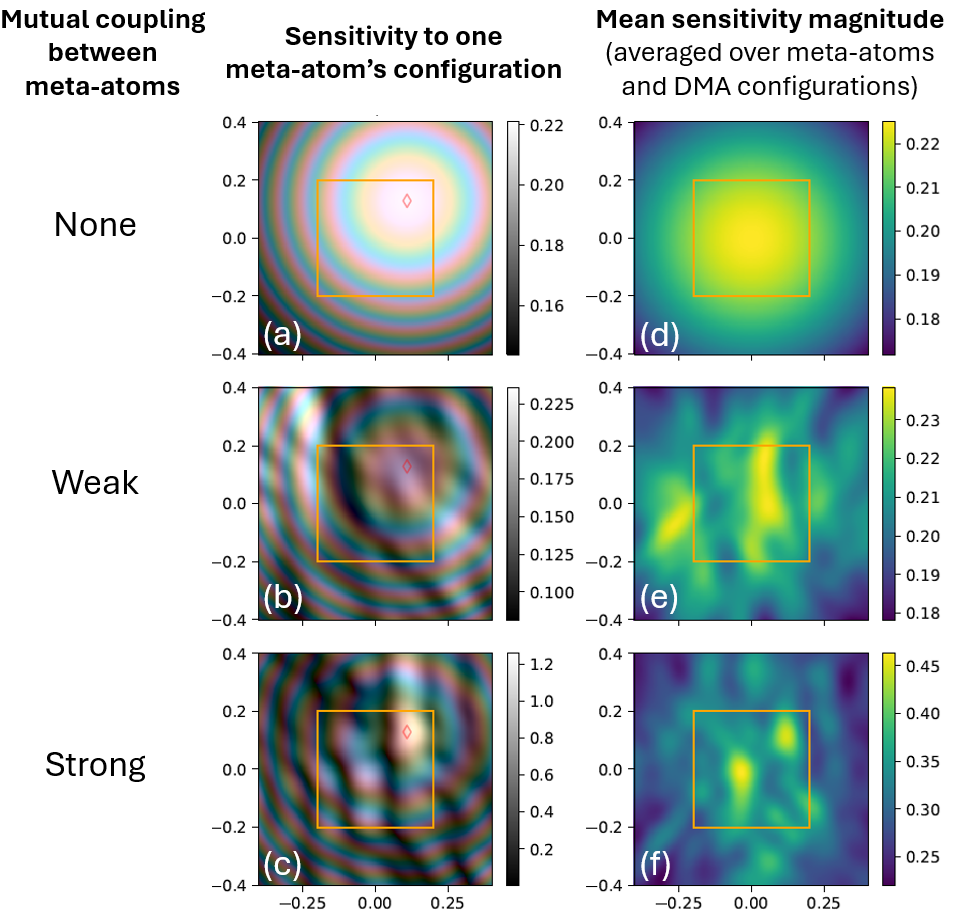}
\caption{\textbf{Radiation pattern sensitivity to DMA configuration for different strengths of mutual coupling between meta-atoms.} The mutual coupling strength between the meta-atoms is controlled by the number of vias (see Fig.~\ref{fig1}). The scenario without mutual coupling is obtained by enforcing a ``unilateral'' approximation under which all couplings except for those from feed to meta-atoms are set to zero. 
(a-c) Radiation pattern sensitivity (partial derivative) with respect to one meta-atom's configuration. Magnitude and phase of the complex-valued sensitivity are represented by luminance and hue, respectively. The considered meta-atom is indicated. (d-f) Average of the sensitivity magnitude over all meta-atoms and for 1000 different random DMA configurations. }
\label{fig2}
\end{figure}

First, we examine the sensitivity of the radiation pattern with respect to the configuration of a single meta-atom for the same uniform DMA configuration as in Fig.~\ref{fig1}(b). We depict the obtained sensitivity in the ROI for three different regimes of mutual coupling strength in Fig.~\ref{fig2}(a-c). As mentioned earlier, we can conveniently vary the mutual coupling strength from weak to strong by adjusting the number of vias. In addition, we also consider the idealistic benchmark case of zero mutual coupling by enforcing a so-called ``unilateral approximation''~\cite{ivrlavc2010toward} consisting in setting all Green's functions to zero except for those from the feed to the meta-atoms. To ensure a fair comparison between different levels of mutual coupling strength, we consider normalized radiation patterns in our analysis such that any potential dependence of the overall level of radiated energy on the mutual coupling strength does \textit{not} impact the reported sensitivity values.

We observe in Fig.~\ref{fig2}(a) that the sensitivity in the ROI resembles the well-known radiation pattern of a single point source in the case of zero mutual coupling. This makes sense because an infinitesimal change of the considered meta-atom's configuration only impacts that meta-atom's dipole moment but not the dipole moments of the other meta-atoms when mutual coupling is completely absent. In contrast, as seen in Fig.~\ref{fig2}(b), even for a weak level of mutual coupling the sensitivity deviates from the point-source radiation pattern, featuring notable interference effects that are particularly visible in terms of its magnitude. Indeed, in the presence of mutual coupling, an infinitesimal change of one meta-atom's polarizability impacts the dipole moments of all meta-atoms, and hence the sensitivity can be interpreted as a superposition of the radiation patterns of as many point sources as there are meta-atoms. Yet, in the weak mutual coupling regime only the infinitesimally reconfigured meta-atom's dipole moment changes substantially, such that the sensitivity map resembles the superposition of multiple point source of which one is ``dominant''. In contrast, in the strong mutual coupling regime considered in Fig.~\ref{fig2}(c), many meta-atoms' dipole moments change substantially upon an infinitesimal change of the considered meta-atom's configuration, resulting in a sensitivity map with speckle-like features that resembles the interference pattern originating from a multitude of significant point sources. Importantly, we note that the sensitivity magnitudes in Fig.~\ref{fig2}(c) are substantially higher than in Fig.~\ref{fig2}(a,b).

Next, we examine the average of the magnitude of the sensitivity map in the ROI over all meta-atoms and 1000 different random DMA configurations. The corresponding results in Fig.~\ref{fig2}(d-f) display a substantial dependence of the sensitivity magnitude on the mutual coupling strength -- note the differences in the utilized colorbar scales. 

\begin{figure}[b]
\centering
\includegraphics [width = 0.6\linewidth]{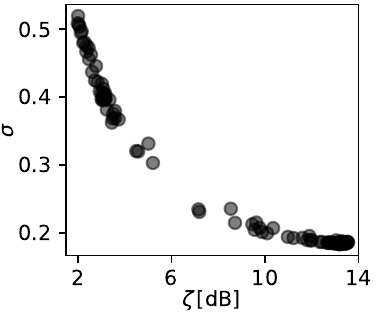}
\caption{\textbf{Average magnitude of the radiation pattern sensitivity $\sigma$ vs. average accuracy $\zeta$ of a linear forward model.} 
$\sigma$ is evaluated based on the partial derivative of the normalized radiation pattern with respect to the configuration of a given meta-atom, averaged over all meta-atoms, 1000 random DMA configurations, and 12 DMA topologies. $\zeta$ is evaluated similarly to a signal-to-noise ratio but using the prediction error of a multi-variable linear regression as ``noise'' (see Sec.~II in~\cite{rabault2024tacit} for technical details), averaged over 12 DMA topologies.}
\label{fig3}
\end{figure}

Incidentally, a similar trend has been previously observed for the conceptually related question of how multi-path propagation impacts the achievable accuracy in wireless localization~\cite{del2021deeply}. Despite clear differences in the considered physical setups, the fundamental wave propagation concepts at play are very related. Whereas the correlation between multiple-scattering strength and wave field sensitivity was found to be favorable for sensing in~\cite{del2021deeply}, in the present article we establish its benefits for wavefront control with DMAs. In both cases, however, stronger multiple scattering also brings about a substantial increase in the complexity of the required signal processing because the non-linearity of the mapping from system configuration to resulting wave field simultaneously increases with the mutual coupling strength, as we now go on to evidence in Fig.~\ref{fig3}.

\begin{figure*}[t]
\centering
\includegraphics [width = 0.9\linewidth]{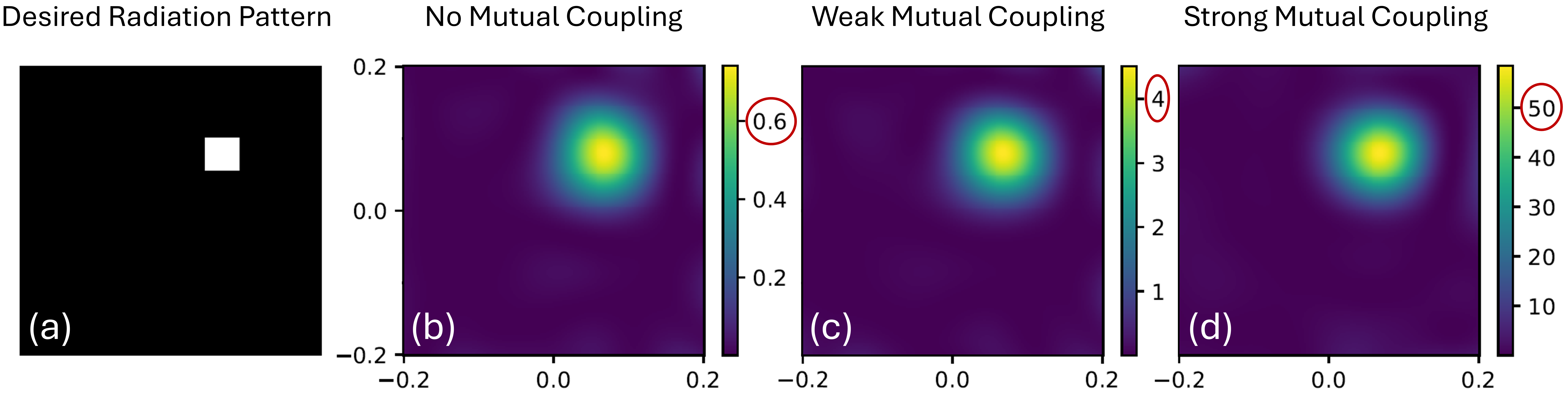}
\caption{\textbf{Radiation pattern synthesis for prototypical beamforming problem as a function of mutual coupling strength. } The targeted radiation pattern is displayed in (a). The normalized radiation patterns obtained with an optimized DMA configuration are displayed for (a) no, (b) weak and (c) strong mutual coupling between the meta-atoms. The three considered DMAs in (b-d) are the same as in Fig.~\ref{fig2}. Note the drastically different colorbar scales in (b-d).}
\label{fig4}
\end{figure*}

Specifically, we now systematically study the trade-off between the favorable increase in sensitivity with increased mutual coupling strength evidenced in Fig.~\ref{fig2} and the unfavorable simultaneous decrease in the predictability of the radiation pattern with a simple linear model. As mentioned earlier, the two trends are, respectively, the ``friend'' side and the ``foe'' side of the same coin that is mutual coupling between the DMA's meta-atoms. To quantify the linear predictability of the radiation pattern, we perform a multi-variable linear regression to obtain the best linear model that maps DMA configuration to radiation pattern. We then determine this linear model's prediction error and evaluate a linearity metric $\zeta$ that is defined analogously to a signal-to-noise ratio except that it uses the prediction error as noise -- technical details on $\zeta$ can be found in Sec.~II of~\cite{rabault2024tacit}. We average the mean sensitivity magnitude $\sigma$ and the linearity metric $\zeta$ over 12 different DMA topologies. Our results are displayed in Fig.~\ref{fig3} and evidence a very clear trade-off. As explained earlier, this trade-off makes perfect sense upon interpreting the mutual coupling strength in terms of a physical multi-bounce picture.

\section{Radiation Pattern Synthesis \\with Optimized DMA Configurations}
\label{sec_synthesis}

Having established that stronger mutual coupling between the DMA's meta-atoms results in a stronger sensitivity of the radiation pattern to the DMA configuration, we now examine how this increased sensitivity manifests itself in an enhanced ability to synthesize a desired radiation pattern by optimizing the DMA configuration. Given the implementation of the differentiable physical forward model outlined in Sec.~\ref{sec_forwardModel} in an automatic differentiation framework, the identification of an optimized DMA configuration can be conveniently performed via error backpropagation (also known as ``adjoint-based'' optimization) by declaring the DMA configuration as trainable weights and defining a cost function that compares the current radiation pattern to the desired one~\cite{del2020learned}. As before, we work with normalized radiation patterns. We consider the prototypical optimization objective from dynamic beamforming displayed in Fig.~\ref{fig4}(a) which requires the maximization of power radiated into a specified direction and a simultaneous minimization of the power radiated into all other directions.

We observe that in all three cases the optimized configuration yields a radiation pattern that conforms well with the objective upon visual inspection -- see Fig.~\ref{fig4}(b-d). Given the relatively simple desired pattern and the large number of continuously tunable meta-atoms, the objective is apparently easy to satisfy in principle. Nonetheless, a drastic difference is apparent in the colorbar scales in Fig.~\ref{fig4}(b-d) which differ by an order of magnitude between each other. Recall that this difference is purely due to the enhanced sensitivity which enables a stronger constructive interference and not due to differences in the overall level of radiated energy thanks to our normalization. Two orders of magnitude difference between no and strong mutual coupling would make a very substantial difference in the signal strength received at the user equipment, and ultimately in the achievable information transfer rate. To summarize, the substantially larger sensitivity under strong mutual coupling manifests itself in a notably improved ability to synthesize favorable radiation patterns, as demonstrated here for a prototypical beamforming problem.

\section{Challenges, Open Questions, \\and Research Directions}
\label{sec_challenges}

Reaping the previously overlooked benefits of mutual coupling in DMAs that we evidenced in this paper promises substantially more control over the radiation pattern with a given number of tunable elements and DMA size. However, these advantages come at the cost of requiring fundamentally different principles for the design and deployment of DMAs. In terms of DMA design, shifting from mitigating mutual coupling to determining what mutual coupling constellations are most beneficial will result in fundamentally different design principles for DMA hardware. In terms of DMA deployment, the strong non-linearity of the forward mapping from DMA configuration to radiation pattern in the strong mutual coupling regime will result in fundamentally different principles underpinning algorithmic DMA software. Specifically, due to this non-linearity, even compactly representing and characterizing the design space for an experimentally given DMA prototype is an open challenge, let alone efficiently exploring it for optimization. 
In the following, we list some of the most important related research challenges to date.

\subsection{Frugal calibration of compact forward models} 

A (preferably differentiable) forward model that accurately captures the non-linear mapping from DMA configuration to radiation pattern for an experimentally given DMA prototype is a crucial prerequisite for optimizing the DMA configuration to achieve diverse wireless communications functionalities. Learning a physics-agnostic forward model appears possible in principle, but already the required thousands (or more) radiation-pattern measurements (each taking multiple minutes) to train an artificial neural network appear prohibitively costly. We expect that purely physics-based models will prevail thanks to their orders-of-magnitude benefits in accuracy, compactness (i.e., number of parameters) and calibration requirements (i.e., number of required calibration examples) that all originate from their \textit{favorable inductive bias}. In the related area of smart radio environments (SREs), these benefits of physical over physics-agnostic models were recently evidenced experimentally~\cite{sol2024experimentally}; one striking consequence of the physics inductive bias was the achievement of high-fidelity model calibration purely based on non-coherent measurements~\cite{sol2024experimentally}. However, for DMAs with strong mutual coupling, corresponding results are missing to date. 

Existing efforts on formulating physics-compliant CCB-DMA models like~\cite{yoo2020analytic} seek to explicitly describe each detail of the DMA architecture in closed form; while this enables numerical studies like the one presented in this article, this approach is very cumbersome and vulnerable to fabrication inaccuracies. We expect that more compact physical models will prevail that only describe the primary entities (feeds and meta-atoms) explicitly while the effects of the other entities (vias, conducting plates, dielectric layers etc.) are lumped into the background Green's functions between the primary entities. This approach, recently established for SREs in~\cite{sol2024experimentally}, substantially reduces the number of parameters, thereby avoiding over-parametrization, and provides robustness against fabrication inaccuracies and other unknowns. Importantly, it can be applied irrespective of the detailed DMA architecture and without knowledge of the latter (maybe except for the meta-atoms' locations).

We emphasize that despite the similarities with SREs, there are important structural differences between physics-compliant models of SREs and DMAs that are to date not well documented. Indeed, measurements in SREs are taken at antennas rather than directly at the reconfigurable entities (e.g., elements of a reconfigurable intelligent surface); in contrast, the radiation pattern of a DMA is a direct linear projection of the tunable meta-atom's dipole moments. In addition, it remains to be determined whether for practical DMAs a dipole approximation of the meta-atoms is sufficient, or whether higher-order truncated multi-pole approximations are required to achieve satisfactory accuracy. A further open question is how to compactly capture frequency selectivity, as required in broadband deployment scenarios.

\subsection{Efficient optimization}

Given a calibrated forward model, efficiently exploring the design space to find optimized DMA configurations is a challenging optimization problem because of its non-linearity, high dimensionality and non-convexity. Detailed comparative studies on optimization strategies based on physics-compliant models for the strong mutual coupling regime are missing to date. An important consideration are restrictions of the meta-atoms' configurations to a discrete set of accessible states; while the backpropagation method can cope with such constraints~\cite{del2020learned} it remains unclear if this yields the best outcomes in the most efficient manner. For instance, coordinate-descent approaches may benefit from rapid forward evaluations despite the matrix inversion at the core of any physical forward model thanks to the Woodbury identity, as developed for SREs in~\cite{prod2023efficient}. Ultimately, it also remains unclear whether optimizations are feasible on the fly, or whether codebook-based approaches will prevail, in which case the computational cost of the optimization algorithms is less critical. 

An optimal DMA configuration should not only realize a desired radiation pattern, but simultaneously minimize the feed's return loss to avoid reflected-power echos and to optimally use the signal power. This multi-objective optimization closely relates to the concept of ``reflectionless signal routing'' for which a rigorous scattering theory is currently under development~\cite{faul2024agile}.

\subsection{Electromagnetic and physics-consistent information-theoretic performance bounds}

Bounds on achievable performances with DMAs, in particular DMAs with mutual coupling, remain an uncharted area to date. This applies both to electromagnetic bounds on achievable radiation patterns, as well as to physics-consistent information-theoretic bounds on wireless communications with DMAs. We expect that physical DMA models will play a crucial role in the development of such bounds because they incorporate the mathematical structure of the governing physical equations. Understanding such bounds will play a pivotal role in guiding the DMA design in terms of key parameters such as the DMA's aperture size, the number of tunable meta-atoms, and the quantization of the accessible meta-atom states.

\subsection{End-to-end hardware design}

Our discussion above of open challenges related to the formulation and calibration of physical DMA models, as well as their use for optimization and bounding, was agnostic to the detailed DMA architecture. However, it is also important to reflect on what DMA hardware designs optimally reap the benefits of mutual coupling. For specific-purpose DMAs, an end-to-end optimization of the hardware layout with the specific purpose in mind may be beneficial. For general-purpose DMAs desired to be capable of synthesizing arbitrarily complex radiation patterns, we hypothesize that no unique optimal hardware layout will exist, but that instead any embodiment favoring strong all-to-all coupling between the primary entities performs well. This would alleviate the hardware design burden because any ergodic chaotic cavity would satisfy this criterion, in line with related recent experiments on agile free-form filters and reflectionless signal routers~\cite{faul2024agile}. Finally, we expect future DMA generations to include multiple feeds, raising design questions about how to optimally place the feeds inside a chaotic cavity (e.g., regarding their spacing).

\section{Conclusion}

DMAs are poised to play an important role in 6G base stations and access points due to their significant benefits in cost and power consumption compared to conventional antenna arrays. Mutual coupling in DMAs is to date treated as a vexing nuance that is ideally mitigated.
In this article, we challenged this viewpoint by establishing an overlooked benefit of mutual coupling between meta-atoms in DMAs: The stronger the meta-atoms are coupled to each other, the more control over the radiation pattern they achieve. Intuitively, we explained this fact in terms of a clear multi-bounce physical picture wherein the wave accumulates more sensitivity to the DMA configuration the more it bounces back and forth between the meta-atoms.
We rigorously evidenced the benefit of mutual coupling between meta-atoms in terms of a physics-compliant analysis of the radiation pattern sensitivity and by considering a concrete pattern synthesis application. However, reaping the unveiled benefit of mutual coupling in DMAs requires fundamentally different design and deployment principles for DMAs. We discussed a set of key open research questions which we expect to pave the way to maximally flexible and compact DMAs by tapping into the potential of mutual coupling between meta-atoms.

\bibliographystyle{IEEEtran}
%\bibliography{references}


\end{document}